\documentclass[aps,prl,twocolumn,groupedaddress]{revtex4}
\usepackage{amsmath}
\usepackage{graphicx}
\usepackage{epsfig}
\begin{document}

\title{Quantum interference in bosonic and fermionic matter-wave amplification}


\author{H. Deng$^{1}$}
\author{Y. Yamamoto$^{1,2}$}
\affiliation{$^{1}$Quantum Entanglement Project, ICORP, JST\\Edward L. Ginzton Laboratory, Stanford University,
Stanford, CA 94305, USA\\$^2$NTT Basic Research Laboratories, Astugishi, Kanagawa, Japan}



\begin{abstract}
We investigate the quantum interference effects in two types of
matter-wave mixing experiments: one with initial matter waves
prepared in independent Fock states (type I) and the other with
each individual particle prepared in a same coherent superposition
of states (type II). In the type I experiment, a symmetric
wavefunction of bosons leads to constructive quantum interference
and shows final state stimulation, while an anti-symmetric
wavefunction of fermions results in destructive quantum
interference and inhibited matter wave mixing. In the type II
experiment, a coherent superposition state leads to constructive
quantum interference and enhanced matter wave mixing for both
bosons and fermions, independent of their quantum statistics.
\end{abstract}


\maketitle

With the realization of Bose-Einstein condensation (BEC) in atoms,
bosonic final state stimulation involving atom condensates has
been studied in superradiance of atoms \cite{Inouye_Sci99} ,
four-wave mixing (FWM)\cite{DengL_NA99} , and matter wave
amplification\cite{Inouye_NA99,Kozuma_Sci99}. Following these
work, it was pointed out that these phenomena are not unique in
boson systems but also possible in fermion
systems\cite{Meystre_PRL01,Inouye_PRL01}. In these experiments
(\cite{Inouye_Sci99}-\cite{Kozuma_Sci99}), the input matter waves,
characterized by their momenta, are all prepared from a same
condensate by a coherent partition process of each individual
particle. Hence the observed nonlinearity can be understood as
collective enhancement effect, analogous to Dicke super-radiance
in an ensemble of two-level atoms, and does not depend on the
quantum statistics of the particles. We call such experiments
\textit{type II} in this paper. In another kind of experiment,
which we call \textit{type I} here, all input matter waves consist
of independent real populations of the particle. In this case,
final state stimulation occurs in a boson system, while inhibition
of matter-wave mixing is expected in a fermion system. Type I
experiment has not yet been performed with atomic BECs, but has
been recently demonstrated with exciton-polaritons in
semiconductors\cite{Robin_PRB00,Robin_PRB02}.

We investigate the two types of matter-wave mixing in terms of the
quantum interference among different paths, naturally originating
from the symmetrization (anti-symmetrization) procedure for type
I, or artificially created from the coherent superposition state
for type II. Figure~1 shows a model FWM experiment where two input
states $|\phi\rangle$ and $|\psi\rangle$ elastically scatter into
two output states $|u\rangle$ and $|v\rangle$ via the two possible
processes (A) and (B), with scattering amplitudes $S_A$ and $S_B$,
respectively\cite{Sakurai_MQM}. The scattering processes are
governed by a unitary operator $\hat{U}_{int}$, s.t.,
\begin{equation}
S_A = \textrm{ }_2\!\,\langle u| _1\!\,\langle v|\hat{U}_{int}(t)
|\phi\rangle _1 |\psi\rangle _2 = \textrm{ }_2\!\,\langle v|
_1\!\,\langle u|\hat{U}_{int}(t) |\psi\rangle _1 |\phi\rangle _2 ,
\end{equation}
\begin{equation}
S_B = \textrm{ }_2\!\,\langle v| _1\!\,\langle u|\hat{U}_{int}(t)
|\phi\rangle _1 |\psi\rangle _2 = \textrm{ }_2\!\,\langle u|
_1\!\,\langle v|\hat{U}_{int}(t) |\psi\rangle _1 |\phi\rangle _2 .
\end{equation}

\begin{figure}[b]
\includegraphics[width=2.5in,angle=-90]{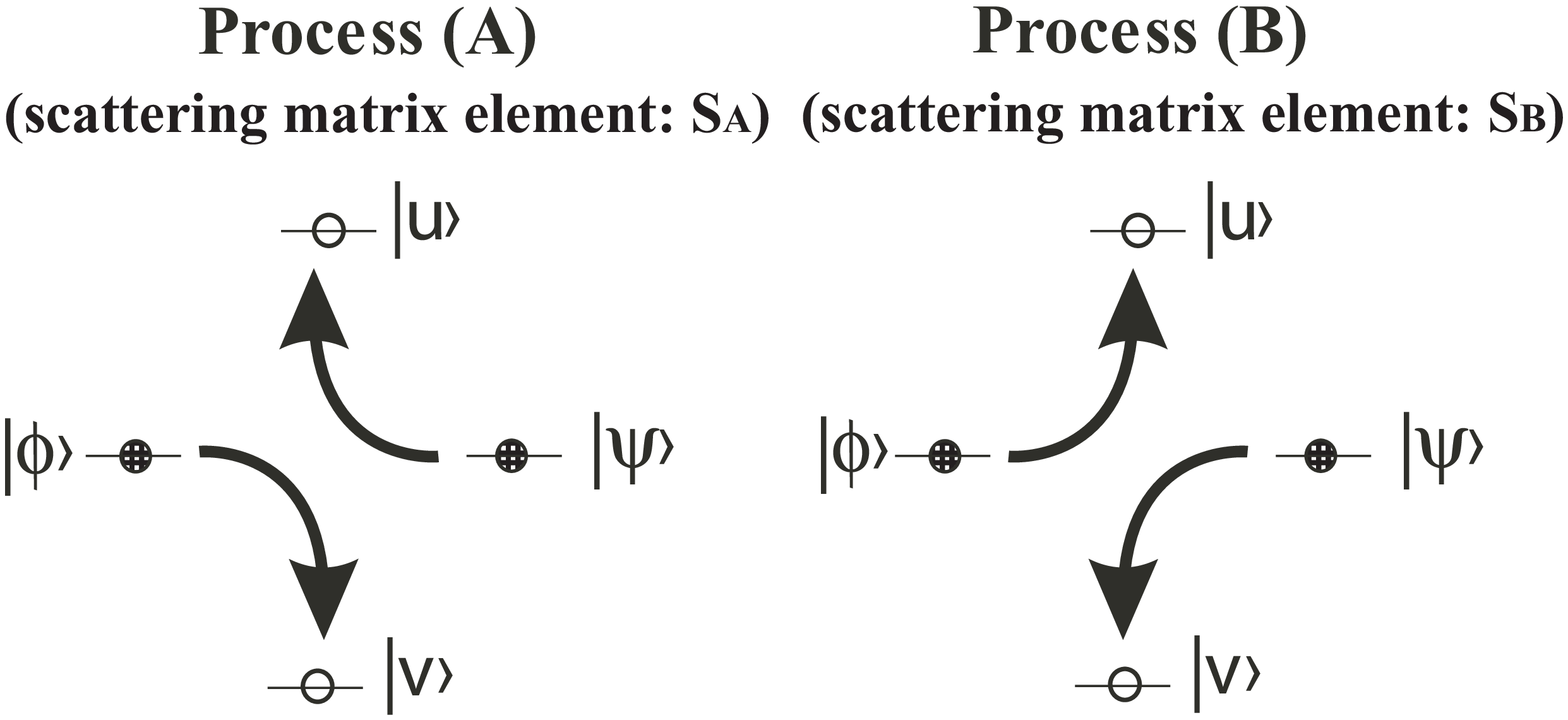}
\caption{} Illustration of the two scattering processes (A) and
(B) in FWM experiments.
\end{figure}

\textbf{Type I experiment} In type I experiment, the four initial
matter waves have definite populations (Fock states). Let's first
consider the case where there is one particle in each of the two
input states and one particle in one output state, with the
initial state of the system:
\begin{align}
|i_I\rangle = & \sum \limits_{i\neq j}\hat{P_{ij}}[ |\phi\rangle
_1 |v\rangle _2 |\psi\rangle _3
] \nonumber \\
 =& \frac{1}{\sqrt{6}}[|\phi\rangle _1 |v\rangle _2
|\psi\rangle _3 \pm |v\rangle _1 |\phi\rangle _2 |\psi\rangle _3
\pm |\psi\rangle _1 |v\rangle _2 |\phi\rangle _3 \nonumber \\
& + |v\rangle _1 |\psi\rangle _2 |\phi\rangle _3 \pm |\phi\rangle
_1 |\psi\rangle _2 |v\rangle _3 + |\psi\rangle _1 |\phi\rangle _2
|v\rangle _3 ], \label{iI}
\end{align}
where $\hat{P}_{ij}$ is the symmetrization or anti-symmetrization
operators\cite{Sakurai_MQM}. The upper sign is for bosons, and the
lower sign is for fermions, in accordance with symmetrization
postulate for bosons and fermions, respectively. Scattering
results in final states with two particles in state $| v\rangle$
and one particle in state $|u\rangle$. Take a final state $|
v\rangle_1 | v\rangle_2 |u\rangle_3$ as an example (Fig.~2(a)), it
can be reached by scattering (A) or (B) from each of the first
four terms in the initial state $|i_I\rangle$. The corresponding
scattering amplitudes for the four paths are $S_A C$, $\pm S_A C$,
$\pm S_B C$, and $S_B C$, where $C$ is a real normalization
factor, $C=1/\sqrt{6}$ in this example. Thus the total scattering
amplitude for bosons adds up to $2(S_A + S_B )C$ as a result of
constructive quantum interference; while for fermions it is
suppressed to zero due to destructive quantum interference. This
illustrates how quantum interference leads to final state
stimulation for bosons, and inhibited FWM for fermions.

\begin{figure}
\includegraphics[width=3.5in]{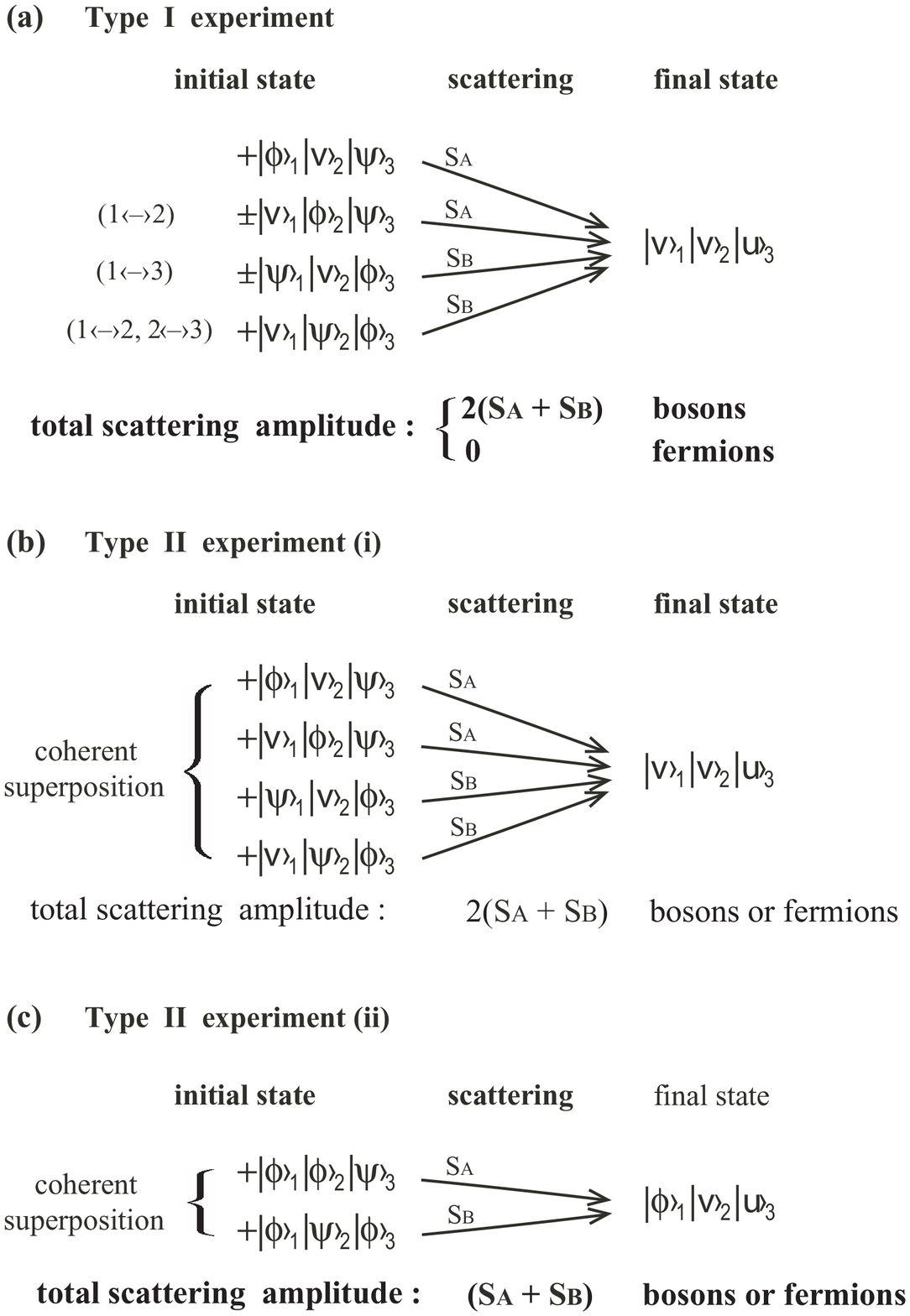}
\caption{} Possible paths of scattering leading to the final state
(a) $|v\rangle _1 |v\rangle _2 |u\rangle _3$ in type I experiment,
(b) $|v\rangle _1 |v\rangle _2 |u\rangle _3$ in type II
experiment, and (c) $|\phi\rangle _1 |v\rangle _2 |u\rangle _3$ in
type II experiment. In (a), the upper sign is for bosons, the
lower sign is for fermions. The total scattering amplitudes is
enhanced for bosons, and is suppressed to zero for fermions in
this type I experiment. In (b) and (c), the scattering amplitudes
are the same for both bosons and fermions. It is enhanced in (c),
but there is no interference leading to enhanced or suppressed
scattering in (c).
\end{figure}

In general, if the initial matter waves of a boson system consists
of $n_1$ particles in $|\phi\rangle$, $n_2$ particles in
$|\psi\rangle$, $n_3$ particles in $| v\rangle$, and $n=n_1 + n_2
+ n_3$ particles in total, the initial-state is:
\begin{align}
|i\rangle _{IB} = & \frac{1}{\sqrt{N}} [
\prod\limits_{i=1}^{n_{1}}|\phi\rangle _i \prod\limits_{j=n_1
+1}^{n_1 + n_2}|\psi\rangle _j
\prod\limits_{k=n_1 + n_2 +1}^{n}| v\rangle _k \nonumber \\
 & + \textrm{ permutation terms due to} \nonumber \\
 &   \textrm{   symmetrization postulate } ].
\label{Ib_i}
\end{align}
All the permutations add to a total number of $N=(^n
_{n_1})(^{n-n_1 } _{n_2})$ different terms in the bracket. Any one
of the $n_1$ particles in $|\phi\rangle$ and any one of the $n_2$
particles in $|\psi\rangle$ can scatter into $| v\rangle$ and
$|u\rangle$ via the two processes (A) and (B). Process (A) results
in a total of $N_2 = N(^{n_1}_1)(^{n_2}_1)$ terms in the final
state $|\tilde{f}\rangle _{IB} = \hat{U}_{int} |i\rangle _{IB}$.
Each of these terms has one particle in $|u\rangle$, $n_1 -1$
particles in $|\phi\rangle$, $n_2 -1$ particles in $|\psi\rangle$,
and $n_3 +1$ particles in $| v\rangle$. Hence there are only $N_3
=(^{n}_{1})(^{n-1}_{n_1 -1})(^{n-n_1}_{n_2 -1})$ physically
distinct terms in the final state. Due to the symmetry property of
the initial state (\ref{Ib_i}), all these $N_3$ terms have the
identical probability amplitude $S_A c_{IB} = S_A
(\frac{N_2}{N_3})/\sqrt{N}$. Same analysis applies to the process
(B) except that the probability amplitude of each term is $S_B
c_{IB} = S_B (\frac{N_2}{N_3})/\sqrt{N}$. The normalized final
state is $|f\rangle_{IB} =
|\tilde{f}\rangle_{IB}/\sqrt{_{IB}\langle\tilde{f}|\tilde{f}\rangle_{IB}}$.
The scattering amplitude is:
\begin{align}
a_{IB} = & _{IB}\langle f|\hat{U}_{int}|i\rangle_{IB} =
_{IB}\langle
f|\tilde{f}\rangle _{IB} = \sqrt{_{IB}\langle\tilde{f}|\tilde{f}\rangle_{IB}} \nonumber\\
=& \sqrt{N_3 c_{IB}^2 } |S_A + S_B |  \nonumber \\
=& \sqrt{n_1 n_2 (n_3 +1)}|S_A + S_B |.
\end{align}
It shows that the scattering rate is proportional to the product
of the numbers of particles ($n_1$ and $n_2$) in the input states
and is enhanced by the initial occupancy $n_3$ of the output state
$| v\rangle$.

In the case of a fermion system, the initial state is:
\begin{widetext}
\begin{align}
|i\rangle _{IF} = & \frac{1}{\sqrt{n!}} [ |\phi^{(1)}\rangle _1
|\phi^{(2)}\rangle _{2} ... |\phi^{(n_1)}\rangle _{n_1}
|\psi^{(1)}\rangle _{n_1 +1}...|\psi^{(n_2 )}\rangle _{n_1 +n_2 }
|v^{(1)}\rangle _{n_1 +n_2 +1}...| v^{(n_3 )}\rangle _n \nonumber \\
 & + \textrm{ permutation terms due to anti-symmetrization postulate} ],
\label{If_i}
\end{align}
\end{widetext}
Here the superscripts $(1), (2),...$ are labels of another quantum
number q, which is conserved during the scattering processes.
Thus, for example, the $n_1$ particles occupy nearly degenerate
but distinct states $|\phi^{(1)}\rangle _1$, $|\phi^{(2)}\rangle
_{2}$, etc. There are a total of $n!$ physically distinct terms in
the bracket. Due to the anti-symmetrized form of the initial
state, scattering amplitude cancel out exactly between different
paths if they lead to a final state with two particles in a same
state $|v\rangle$. For a scattering process in which two particles
are scattered into initially unoccupied $|v\rangle$ and
$|u\rangle$, there is no other paths interfering with it.
Explicitly, the total scattering amplitude $ _{IF}\langle
f|\hat{U}_{int}|i\rangle_{IF}$ is:
\begin{widetext}
\begin{equation}
a_{IF}=\left\{\begin{array}{ll} 0, &\textrm{if } n_3 \geq n_1 ,n_2 ,\\
\sqrt{(n_1 -n_3)n_2 }|S_A |, &\textrm{if } n_1 >n_3 \geq n_2 ,\\
\sqrt{(n_2 -n_3)n_1 }|S_B |, &\textrm{if } n_2 >n_3 \geq n_1 ,\\
\textrm{}[(n_1 -n_3)n_2 |S_A |^2 + (n_2 -n_3)n_1 |S_B |^2
+ 2(n_2 -n_3 )(S_A S_B ^* + S_A ^* S_B )]^{1/2}, & \textrm{if } n_1 > n_2 >n_3 ,\\
\textrm{}[(n_1 -n_3)n_2 |S_A |^2 + (n_2 -n_3)n_1 |S_B |^2  + 2(n_1
-n_3 )(S_A S_B ^* + S_A ^* S_B )]^{1/2}, & \textrm{if } n_2
> n_1 >n_3 .
\end{array}
\right.
\end{equation}
\end{widetext}
It shows that if there are initially more particles in an ensemble
of nearly degenerate states $| v\rangle$ than in $|\phi\rangle$
and $|\psi\rangle$, the scattering into $| v\rangle$ and
$|u\rangle$ is completely suppressed. Otherwise, the amplitude is
non-zero but still suppressed by the increase of $n_3$.

\textbf{Type II experiment} In contrast to the type I experiment,
each particle in the initial matter waves of a type II experiment
is in a same coherent superposition of the states $|\phi\rangle$,
$|\psi\rangle$, and $| v\rangle$. In a boson system, if there are
n particles in total, and each particle prepared in an identical
superposition state,
the initial-state wavefunction of bosons is:
\begin{equation}
|i\rangle _{IIB}  = \prod \limits_{i=1}^{n}
 (\sqrt{\frac{1-\epsilon}{2}}|\phi\rangle_i +
\sqrt{\frac{1-\epsilon}{2}}|\psi\rangle_i +
\sqrt{\epsilon}| v\rangle_i ) 
\label{IIB_i}
\end{equation}
The expansion of (\ref{IIB_i}) consists of a total of $3^n$
different terms. There are $N_{mk}=(^n _m)(^{n-m} _k)$ terms which
have m particles in $|\phi\rangle$, k particles in $|\psi\rangle$
and (n-m-k) particles in $| v\rangle$. Here m takes values from 0
to n, for each m, k takes values from 0 to n-m. We call this group
of $N_{mk}$ terms as (m,k) group. All terms in the same (m,k)
group have the same probability amplitude $c_{mk}^{0}=
(\frac{1-\epsilon}{2})^{n/2} \eta^{\frac{n-m-k}{2}}$, where
$\eta=\frac{2\epsilon}{1-\epsilon}$. It is obvious from the
expansion that $|i\rangle _{IIB}$ is already fully symmetric, and
no additional symmetrization procedure is necessary.

For each (m,k) group, possible scattering of a pair of
$|\phi\rangle _i$ and $|\psi\rangle _j$ into $| v\rangle _i
|u\rangle _j$ via process (A) results in a total of $N_2 '= mk
N_{mk}$ terms, each of which has one particle in $|u\rangle$,
(m-1) particles in $|\phi\rangle$, (k-1) particles in
$|\psi\rangle$ and (n-m-k+1) particles in $| v\rangle$. However,
there are only $N_3 ' =(^n _1)(^{n-1}_{m-1})(^{n-m}_{k-1})$
physically distinct terms. Since the initial state is symmetric,
the initial group (m,k) is scattered into $N_3 '$ different terms,
all with the same probability amplitude $S_A c_{mk} ' = S_A
c_{mk}^0 \frac{N_2 '}{N_3 '} = c_{mk}^0\cdot 2(n-m-k+1)$.
Similarly, scattering via process (B) contributes $S_B c_{mk}'$ to
the probability amplitude. The final state
$|\tilde{f}\rangle_{IIB}=\hat{U}_{int}|i\rangle_{IIB}$ is a sum of
terms scattered from all (m,k) groups. Hence the total scattering
amplitude is:
\begin{align}
a_{IIB} = &\sqrt{_{IIB}\langle\tilde{f}|\tilde{f}\rangle_{IIB}}\nonumber \\
=&\sqrt{\sum\limits_{m=1}^{n-1}\sum\limits_{k=1}^{n-m}
N_3 '(m,k) (c' _{mk}|S_A + S_B |)^2 } \nonumber \\
=&\sqrt{[\frac{1-\epsilon}{2}n][\frac{1-\epsilon}{2}(n-1)][\epsilon
(n-2) +1]}|S_A + S_B | . \label{aIIB}
\end{align}
When $n \gg 1$, the scattering rate is again proportional to the
product of the average numbers of particles in the two input
states (the first two terms in the last line of (\ref{aIIB})
corresponding to $n_1$ and $n_2$ in type I experiment), and is
enhanced by the final state population by a factor
$\epsilon(n-2)+1$, where $\epsilon(n-2)$, the average population
in $|v\rangle$, corresponds to $n_3$ in the type I experiment.

For type II experiment, a fermion system has exactly the same
scattering amplitude. The initial state for fermions is
\begin{widetext}
\begin{align}
|i\rangle _{IIF}  = & \frac{1}{N'}
[(\frac{1-\epsilon}{2})^{n/2}\prod\limits_{i=1}^{n}(|\phi^{(i)}\rangle_i
+ |\psi^{(i)}\rangle_i + \sqrt{\eta}| v^{(i)}\rangle_i ) \nonumber \\
 & + \textrm{ permutation terms due to anti-symmetrization postulate} ].
\label{IIF_i},
\end{align}
\end{widetext}
where $N'$ is a normalization factor. Since each particle has a
different quantum number q, and q is conserved under the operation
of $\hat{U}_{int}$, terms in the expansions of different
anti-symmetrization groups do not interfere with each other, even
after scattering. At the same time, all anti-symmetrization groups
have identical scattering characteristics. Therefore it is
sufficient to consider only the first line of (\ref{IIF_i}),
setting $N'=1$. In another word, symmetrization postulate and thus
quantum statistics does not affect the scattering amplitude in
type II experiment. Moreover, the label for quantum number q in
(\ref{IIF_i}) has a one to one correspondence to the label of the
particle number, so the label for quantum number q can be suppress
and (\ref{IIF_i}) is reduced to the same form as (\ref{IIB_i}),
leading to $a_{IIF} =a_{IIB}$.

We again consider a simple case of three particles, each particle
occupying the three states $|\phi\rangle$, $|\psi\rangle$ and $|
v\rangle$ with equal probability. Then the initial state of bosons
and the reduced initial state of fermions has the same form:
\begin{align*}
|i> = &(|\phi\rangle _1 + |\psi\rangle _1 +| v\rangle _1 )\otimes
(|\phi\rangle _2 + |\psi\rangle _2 +| v\rangle _2 ) \\
&\otimes(|\phi\rangle _3 + |\psi\rangle _3 +| v\rangle _3 )
\end{align*}
To reach a final state $| v\rangle _1 | v\rangle _2 |u\rangle _3$,
there are four possible paths, as illustrated in Fig.~2(b). The
corresponding four terms in the initial state $|i\rangle$
originate from a coherent superposition state instead of the
symmetrization or anti-symmetrization procedure. Therefore paths
from each process are additive for both bosons and fermions. The
constructive interference between different paths leads to
enhanced scattering amplitude. To have an intuitive picture of the
enhancement by the final state occupancy (corresponding to the
$\epsilon(n-2)$ term in (\ref{aIIB})), we consider the case where
there are no particle in $|v\rangle$ before scattering. Then only
two paths are possible, as shown in Fig.~2(c), and no interference
terms to lead to enhancement in this case.

As discussed above, type II experiment will produce identical
enhancement in scattering amplitude for both boson and fermion
systems, given that the initial state of the system is prepared as
a coherent superposition of all three states $|\phi\rangle$,
$|\psi\rangle$ and $|v\rangle$. The enhancement comes from
constructive multi-particle interference, where the different
paths are created by preparing the initial state in a coherent
superposition-state. Type I experiment, however, will reveal final
state stimulation for bosons and inhibited-FWM for fermions (Pauli
blocking). The enhancement and inhibition in this case come from
constructive and destructive multi-particle interference, where
the different paths stem from the symmetrization and
anti-symmetrization postulate. So only type I experiment tests the
true quantum statistics of the system. The final state
stimulation\cite{Robin_PRB00}, matter wave
amplification\cite{Robin_PRB02} and condensation of exciton
polaritons\cite{sci02} have been demonstrated in this type of
experiment, but the counterpart experiments in atomic systems are
yet to be observed.

\bibliography{arxiv0210}

\begin{thebibliography}{10}
\expandafter\ifx\csname natexlab\endcsname\relax\def\natexlab#1{#1}\fi
\expandafter\ifx\csname bibnamefont\endcsname\relax
  \def\bibnamefont#1{#1}\fi
\expandafter\ifx\csname bibfnamefont\endcsname\relax
  \def\bibfnamefont#1{#1}\fi
\expandafter\ifx\csname citenamefont\endcsname\relax
  \def\citenamefont#1{#1}\fi
\expandafter\ifx\csname url\endcsname\relax
  \def\url#1{\texttt{#1}}\fi
\expandafter\ifx\csname urlprefix\endcsname\relax\def\urlprefix{URL }\fi
\providecommand{\bibinfo}[2]{#2}
\providecommand{\eprint}[2][]{\url{#2}}

\bibitem[{\citenamefont{Inouye et~al.}(1999{\natexlab{a}})\citenamefont{Inouye,
  Chikkatur, StamperKurn, Stenger, Pritchard, and Ketterle}}]{Inouye_Sci99}
\bibinfo{author}{\bibfnamefont{S.}~\bibnamefont{Inouye}},
  \bibinfo{author}{\bibfnamefont{A.}~\bibnamefont{Chikkatur}},
  \bibinfo{author}{\bibfnamefont{D.}~\bibnamefont{StamperKurn}},
  \bibinfo{author}{\bibfnamefont{J.}~\bibnamefont{Stenger}},
  \bibinfo{author}{\bibfnamefont{D.}~\bibnamefont{Pritchard}},
  \bibnamefont{and} \bibinfo{author}{\bibfnamefont{W.}~\bibnamefont{Ketterle}},
  \bibinfo{journal}{Science} \textbf{\bibinfo{volume}{285}},
  \bibinfo{pages}{571} (\bibinfo{year}{1999}{\natexlab{a}}).

\bibitem[{\citenamefont{Deng et~al.}(1999)\citenamefont{Deng, Hagley, Wen,
  Trippenbach, Band, Julienne, Simsarian, Helmerson, Rolston, and
  Phillips}}]{DengL_NA99}
\bibinfo{author}{\bibfnamefont{L.}~\bibnamefont{Deng}},
  \bibinfo{author}{\bibfnamefont{E.}~\bibnamefont{Hagley}},
  \bibinfo{author}{\bibfnamefont{J.}~\bibnamefont{Wen}},
  \bibinfo{author}{\bibfnamefont{M.}~\bibnamefont{Trippenbach}},
  \bibinfo{author}{\bibfnamefont{Y.}~\bibnamefont{Band}},
  \bibinfo{author}{\bibfnamefont{P.}~\bibnamefont{Julienne}},
  \bibinfo{author}{\bibfnamefont{J.}~\bibnamefont{Simsarian}},
  \bibinfo{author}{\bibfnamefont{K.}~\bibnamefont{Helmerson}},
  \bibinfo{author}{\bibfnamefont{S.}~\bibnamefont{Rolston}}, \bibnamefont{and}
  \bibinfo{author}{\bibfnamefont{W.}~\bibnamefont{Phillips}},
  \bibinfo{journal}{Nature} \textbf{\bibinfo{volume}{398}},
  \bibinfo{pages}{218} (\bibinfo{year}{1999}).

\bibitem[{\citenamefont{Inouye et~al.}(1999{\natexlab{b}})\citenamefont{Inouye,
  Pfau, Gupta, Chikkatur, Gorlitz, Pritchard, and Ketterle}}]{Inouye_NA99}
\bibinfo{author}{\bibfnamefont{S.}~\bibnamefont{Inouye}},
  \bibinfo{author}{\bibfnamefont{T.}~\bibnamefont{Pfau}},
  \bibinfo{author}{\bibfnamefont{S.}~\bibnamefont{Gupta}},
  \bibinfo{author}{\bibfnamefont{A.}~\bibnamefont{Chikkatur}},
  \bibinfo{author}{\bibfnamefont{A.}~\bibnamefont{Gorlitz}},
  \bibinfo{author}{\bibfnamefont{D.}~\bibnamefont{Pritchard}},
  \bibnamefont{and} \bibinfo{author}{\bibfnamefont{W.}~\bibnamefont{Ketterle}},
  \bibinfo{journal}{Nature} \textbf{\bibinfo{volume}{402}},
  \bibinfo{pages}{641} (\bibinfo{year}{1999}{\natexlab{b}}).

\bibitem[{\citenamefont{Kozuma et~al.}(1999)\citenamefont{Kozuma, Suzuki,
  Torii, Sugiura, Kuga, Hagley, and Deng}}]{Kozuma_Sci99}
\bibinfo{author}{\bibfnamefont{M.}~\bibnamefont{Kozuma}},
  \bibinfo{author}{\bibfnamefont{Y.}~\bibnamefont{Suzuki}},
  \bibinfo{author}{\bibfnamefont{Y.}~\bibnamefont{Torii}},
  \bibinfo{author}{\bibfnamefont{T.}~\bibnamefont{Sugiura}},
  \bibinfo{author}{\bibfnamefont{T.}~\bibnamefont{Kuga}},
  \bibinfo{author}{\bibfnamefont{E.}~\bibnamefont{Hagley}}, \bibnamefont{and}
  \bibinfo{author}{\bibfnamefont{L.}~\bibnamefont{Deng}},
  \bibinfo{journal}{Science} \textbf{\bibinfo{volume}{286}},
  \bibinfo{pages}{2309} (\bibinfo{year}{1999}).

\bibitem[{\citenamefont{Moore and Meystre}(2001)}]{Meystre_PRL01}
\bibinfo{author}{\bibfnamefont{M.}~\bibnamefont{Moore}} \bibnamefont{and}
  \bibinfo{author}{\bibfnamefont{P.}~\bibnamefont{Meystre}},
  \bibinfo{journal}{Phys. Rev. Lett.} \textbf{\bibinfo{volume}{86}},
  \bibinfo{pages}{4199} (\bibinfo{year}{2001}).

\bibitem[{\citenamefont{Ketterle and Inouye}(2001)}]{Inouye_PRL01}
\bibinfo{author}{\bibfnamefont{W.}~\bibnamefont{Ketterle}} \bibnamefont{and}
  \bibinfo{author}{\bibfnamefont{S.}~\bibnamefont{Inouye}},
  \bibinfo{journal}{Phys. Rev. Lett.} \textbf{\bibinfo{volume}{86}},
  \bibinfo{pages}{4203} (\bibinfo{year}{2001}).

\bibitem[{\citenamefont{Huang et~al.}(2000)\citenamefont{Huang, Tassone, and
  Yamamoto}}]{Robin_PRB00}
\bibinfo{author}{\bibfnamefont{R.}~\bibnamefont{Huang}},
  \bibinfo{author}{\bibfnamefont{F.}~\bibnamefont{Tassone}}, \bibnamefont{and}
  \bibinfo{author}{\bibfnamefont{Y.}~\bibnamefont{Yamamoto}},
  \bibinfo{journal}{Phys. Rev. B} \textbf{\bibinfo{volume}{61}},
  \bibinfo{pages}{R7854} (\bibinfo{year}{2000}).

\bibitem[{\citenamefont{Huang et~al.}(2002)\citenamefont{Huang, Yamamoto,
  Andr{\'e}, Bleuse, Muller, and Ulmer-Tuffigo}}]{Robin_PRB02}
\bibinfo{author}{\bibfnamefont{R.}~\bibnamefont{Huang}},
  \bibinfo{author}{\bibfnamefont{Y.}~\bibnamefont{Yamamoto}},
  \bibinfo{author}{\bibfnamefont{R.}~\bibnamefont{Andr{\'e}}},
  \bibinfo{author}{\bibfnamefont{J.}~\bibnamefont{Bleuse}},
  \bibinfo{author}{\bibfnamefont{M.}~\bibnamefont{Muller}}, \bibnamefont{and}
  \bibinfo{author}{\bibfnamefont{H.}~\bibnamefont{Ulmer-Tuffigo}},
  \bibinfo{journal}{Phys. Rev. B} \textbf{\bibinfo{volume}{65}},
  \bibinfo{pages}{165314} (\bibinfo{year}{2002}).

\bibitem[{\citenamefont{Sakurai}(1995)}]{Sakurai_MQM}
\bibinfo{author}{\bibfnamefont{J.~J.} \bibnamefont{Sakurai}},
  \emph{\bibinfo{title}{Modern Quantum Mechanics}}
  (\bibinfo{publisher}{Addison-Welsley}, \bibinfo{address}{New York},
  \bibinfo{year}{1995}).

\bibitem[{\citenamefont{Deng et~al.}(2002)\citenamefont{Deng, Weihs, Santori,
  Bloch, and Yamamoto}}]{sci02}
\bibinfo{author}{\bibfnamefont{H.}~\bibnamefont{Deng}},
  \bibinfo{author}{\bibfnamefont{G.}~\bibnamefont{Weihs}},
  \bibinfo{author}{\bibfnamefont{C.}~\bibnamefont{Santori}},
  \bibinfo{author}{\bibfnamefont{J.}~\bibnamefont{Bloch}}, \bibnamefont{and}
  \bibinfo{author}{\bibfnamefont{Y.}~\bibnamefont{Yamamoto}},
  \bibinfo{journal}{Science} \textbf{\bibinfo{volume}{298}},
  \bibinfo{pages}{199 } (\bibinfo{year}{2002}).

\end{thebibliography}

\end{document}